\documentclass[superscriptaddress,twocolumn,showpacs,preprintnumbers,amsmath,amssymb,nofootinbib,ventriculography,longbibliography]{revtex4-1}

\usepackage{graphicx}
\usepackage[usenames,dvipsnames]{xcolor}

\usepackage{amsmath}
\usepackage{amssymb}
\usepackage{amsfonts}
\usepackage{enumerate}

\newcommand{\eps}{\varepsilon}

\newcommand{\defterm}[1]{\emph{#1}}              

\newcommand{\tens}[1]{{\boldsymbol{#1}}}       
\newcommand{\ts}[1]{{\boldsymbol{#1}}}         

\newcommand{\grad}{{\tens{d}}}                 
\newcommand{\covd}{{\tens{\nabla}}}            

\newcommand{\A}[1]{A^{\!(#1)}}                 

\newcommand{\dg}{{{n}}}                           

\newcommand{\enf}[1]{{\tens{e}^{#1}}}                  
\newcommand{\ehf}[1]{{\hat{\tens{e}}^{#1}}}            
\newcommand{\ezf}{{\hat{\tens{e}}^{0}}}                

\newcommand{\be}{\begin{equation}}             
\newcommand{\ee}{\end{equation}}               
\newcommand{\ba}{\begin{eqnarray}}             
\newcommand{\ea}{\end{eqnarray}}               

\begin{document}

\title{New Metrics Admitting the Principal Killing--Yano Tensor}

\author{Valeri P. Frolov}

\email{frolov@phys.ualberta.ca}

\affiliation{Theoretical Physics Institute, University of Alberta, Edmonton,
Alberta, Canada T6G 2G7}

\author{Pavel Krtou\v{s}}

\email{Pavel.Krtous@utf.mff.cuni.cz}

\affiliation{Institute of Theoretical Physics,
Faculty of Mathematics and Physics, Charles University,
V~Hole\v{s}ovi\v{c}k\'ach~2, Prague, 18000, Czech Republic}

\author{David Kubiz\v n\'ak}

\email{dkubiznak@perimeterinstitute.ca}

\affiliation{Perimeter Institute, 31 Caroline Street North, Waterloo, ON, N2L 2Y5, Canada}


\date{December 21, 2017}

\begin{abstract}

It is believed that in any number of dimensions the off-shell Kerr--NUT--(A)dS metric represents a unique geometry admitting the principal (rank 2, non-degenerate, closed conformal Killing--Yano) tensor. The original proof relied on the Euclidean signature and therein natural assumption that the eigenvalues of the principal tensor have gradients of spacelike character. In this paper we evade this common wisdom and construct new classes of Lorentzian (and other signature) off-shell metrics admitting the principal tensor with `null eigenvalues', uncovering so a much richer structure of spacetimes with principal tensor in four and higher dimensions. A few observations regarding the Kerr--Schild ansatz are also made.
\end{abstract}


\maketitle

\section{Introduction}\label{sec1}

It is a remarkable property of the {\em Kerr--NUT--(A)dS family} of spacetimes \cite{Carter:1968cmp, Chen:2006xh} that they admit a hidden symmetry encoded in the principal Killing--Yano tensor. This is true in four \cite{Penrose:1973, Floyd:1973} as well as in higher dimensions \cite{Kubiznak:2006kt}. Many of the basic characteristics
of this family can be directly linked to this
tensor and derived from its very existence. For example, the principal tensor generates a `Killing tower' of symmetries
that guarantee complete integrability of geodesic motion and separability of the Hamilton--Jacobi, Klein--Gordon, and Dirac equations. The type D property of these spacetimes can easily be derived from the integrability condition for the principal tensor, and its eigenvectors are intrinsically related to the principal null directions and a possibility to cast the metric in the Kerr--Schild form, we refer the interested reader to a recent review \cite{FrolovKrtousKubiznak:2017review}.

The {\em principal tensor} is a non-degenerate, closed conformal Killing--Yano 2-form. {In $D$ number of spacetime dimensions (in what follows we assume $D\geq 4$), it obeys} an equation
\be\label{PKYT}
\nabla_c h_{ab}=g_{ca}\xi_b-g_{cb}\xi_a\,,\quad \xi_a=\frac{1}{D-1}\nabla^b h_{ba}\,,
\ee
or in the language of differential forms
\be\label{defPKY}
\covd_{\!\ts{X}}\ts{h} = \ts{X}\wedge \ts{\xi}\,,\quad \ts{\xi}=\frac{1}{D-1} \covd\cdot\ts{h}\,,
\ee
{where $\ts{X}$ is an arbitrary vector field.}
Since $\ts{h}$ is closed, there exists at least locally a potential 1-form  $\ts{b}$ such that
\be
  \ts{h}=\grad \ts{b}\, .
\ee
The condition of non-degeneracy means that the principal tensor has the maximal possible (matrix) rank and possesses the maximal number of functionally independent eigenvalues.

The {\em classification} of higher-dimensional metrics  admitting the principal tensor has been attempted in \cite{Houri:2007xz, Krtous:2008tb, Houri:2008th, Houri:2008ng}---the off-shell Kerr--NUT--(A)dS metric was found to be a unique solution admitting the principal tensor. As a starting point of the derivation, the authors considered a Riemannian metric $\ts{g}$ in
\be
D=2n+\epsilon\,,
\ee
dimensions, with $\epsilon=0,1$ in even, odd dimensions, and the principal tensor was written in the \defterm{Darboux frame},
\begin{equation}\label{orthonormmtrc}
\begin{gathered}
\ts{h} = \sum_{\mu=1}^n x_\mu \enf{\mu}\wedge\ehf{\mu}\,,\\
\ts{g} = \sum_{\mu=1}^n \bigl(
      \enf\mu\,\enf\mu + \ehf\mu\,\ehf\mu
      \bigr) + \eps\, \ezf\,\ezf\,.
\end{gathered}
\end{equation}
Here, $(\enf{\mu}, \ehf{\mu},\ezf)$ is an orthonormal frame and
the quantities $x_\mu$ are related to the `eigenvalues' of the 2-form $\ts{h}$.
The condition of non-degeneracy of the principal tensor
requires that there are exactly ${\dg}$ non-vanishing and functionally independent eigenvalues ${x_\mu}$ whose gradients are linearly independent.
Obviously, the Euclidean signature is at least formally assumed and the spatial character of the gradients,
\be
(\grad x_\mu)^2>0\,,
\ee
naturally follows.

Provided these assumptions, it was shown in \cite{Houri:2007xz, Krtous:2008tb, Houri:2008th, Houri:2008ng} that the most general metric admitting the principal tensor takes the {\em canonical off-shell Kerr--NUT--(A)dS form}. In even dimensions it reads
\be\label{KerrNUTAdSmetric}
\tens{g}_{2n}
  =\sum_{\mu=1}^\dg\;\biggl[\; \frac{U_\mu}{X_\mu}\,{\grad x_{\mu}^{2}}
  +\, \frac{X_\mu}{U_\mu}\,\Bigl(\,\sum_{j=0}^{\dg-1} \A{j}_{\mu}\grad\psi_j \Bigr)^{\!2}
  \;\biggr]\,.
\ee
In odd dimensions we have two possibilities
\ba
\tens{g}_{2n+1}^{(1)}&=&\tens{g}_{2n}+\frac{c}{\A{\dg}}\Bigl(\sum_{k=0}^\dg \A{k}\grad\psi_k\!\Bigr)^{\!2}\;,\label{Kerrodd}\\
\tens{g}_{2n+1}^{(2)}&=&\tens{g}_{2n}+c\A{n}\grad \psi_n^2\,.\label{Kerreven}
\ea
Here, {$c$ is an arbitrary parameter,} the functions ${\A{k}}$, ${\A{j}_\mu}$, and ${U_\mu}$ are `symmetric polynomials' of coordinates ${x_\mu}$:
\begin{equation}\label{AUdefs}
\begin{gathered}
  \A{k}=\!\!\!\!\!\sum_{\substack{\nu_1,\dots,\nu_k=1\\\nu_1<\dots<\nu_k}}^\dg\!\!\!\!\!x^2_{\nu_1}\dots x^2_{\nu_k}\;,
\quad
  \A{j}_{\mu}=\!\!\!\!\!\sum_{\substack{\nu_1,\dots,\nu_j=1\\\nu_1<\dots<\nu_j\\\nu_i\ne\mu}}^\dg\!\!\!\!\!x^2_{\nu_1}\dots x^2_{\nu_j}\;,\\
  U_{\mu}=\prod_{\substack{\nu=1\\\nu\ne\mu}}^\dg(x_{\nu}^2-x_{\mu}^2)\;,
\end{gathered}
\end{equation}
and each metric function ${X_\mu}$ is a function of a single coordinate ${x_\mu}$:
\begin{equation}\label{Xfcdependence}
    X_\mu=X_\mu(x_\mu)\;.
\end{equation}
To indicate that these functions are not specified, we call such a metric {\em off-shell}.
The Darboux frame \eqref{orthonormmtrc} in these coordinate reads
\begin{gather}
\enf\mu = {\Bigl(\frac{U_\mu}{X_\mu}\Bigr)^{\!\frac12}}\grad x_{\mu}\;,\;\;
\ehf\mu = {\Bigl(\frac{X_\mu}{U_\mu}\Bigr)^{\!\frac12}}
  \sum_{j=0}^{\dg-1}\A{j}_{\mu}\grad\psi_j\;,\;\nonumber\\[-1.5ex]
\label{DarbouxFrameCoor}\\[-1.5ex]
\ezf_{(1)} = {\Bigl(\frac{c}{\A{\dg}}\Bigr)^{\frac12}}\,\sum_{k=0}^{\dg}\A{k}\grad\psi_k\;,\;\;
\ezf_{(2)} = \bigl(c\A{\dg}\bigr)^{\frac12}\,\grad\psi_n\,.\nonumber
\end{gather}

The principal tensor is given by
\begin{equation}\label{PCCKY}
  \ts{h} = \sum_{\mu=1}^{\dg} x_\mu\, \grad x_\mu \wedge
    \Bigl(\sum_{k=0}^{\dg-1}\A{k}_\mu \grad\psi_k\Bigr)
         = \sum_\mu x_\mu\, \enf\mu\wedge\ehf\mu\;,
\end{equation}
and can be derived from the potential
\begin{equation}\label{PCCKYpot}
\ts{b} = \frac12 \sum_{k=0}^{\dg-1}\A{k+1}\grad\psi_k\;.
\end{equation}
This single object generates a whole `Killing tower' of explicit and hidden symmetries, including Killing vectors, rank-2 Killing tensors, and increasing rank Killing--Yano tensors, see \cite{FrolovKrtousKubiznak:2017review}.

When the vacuum Einstein equations are imposed,
\be
R_{ab}-\frac12R g_{ab} + \Lambda g_{ab}=0\,,
\ee
the metric functions take the following on-shell form \citep{Chen:2006xh, Hamamoto:2006zf, Houri:2008th, Houri:2008ng}:
\ba\label{Xmus}
\mbox{even}\ D:&&\quad X_\mu=\sum_{k=0}^n c_k x_\mu^{2k}-2b_\mu x_\mu\,,\nonumber\\
\mbox{odd}\ D:&& \quad X_\mu^{(1)}=\sum_{k=1}^n c_k x_\mu^{2k}-2b_\mu-\frac{c}{x_\mu^2}\,,\\
&&\quad X_\mu^{(2)}=\sum_{k=1}^n c_k x_\mu^{2k}-2b_\mu\,,\nonumber
\ea
where the parameter $c_\dg$ is related to the cosmological constant,
\be\label{cnpar}
\Lambda=\frac{1}{2}(-1)^\dg(D-1)(D-2)c_\dg\,,
\ee
while other parameters $c_k, b_\mu$, and (in odd dimensions) $c$ are related to rotations, mass, and NUT parameters, see \cite{FrolovKrtousKubiznak:2017review} for discussion.

Although this result has been derived assuming the formal Euclidean signature, it has been used also for the Lorentzian case. Indeed, using a suitable Wick rotation (see following sections) and choosing carefully coordinate ranges, the metric \eqref{KerrNUTAdSmetric}--\eqref{Kerreven} with metric functions \eqref{Xmus} can be a Lorentzian solution of Einstein's equations \cite{FrolovKrtousKubiznak:2017review}. For this reason the uniqueness of the metric admitting the principal tensor has been usually formulated regardless of the signature.

However satisfactory this uniqueness result is in the Euclidean signature, in what follows we will show that it is no longer true in the spacetimes of Lorentzian and other signatures. In particular, considering such signatures, the principal tensor may possess null eigenvalues, that is eigenvalues characterized by a null gradient. Such a possibility has not been {considered} 
in the original construction presented in \cite{Houri:2007xz, Krtous:2008tb, Houri:2008th, Houri:2008ng}. This allows one to construct new canonical metric elements with the principal tensor. In this paper we present several such new metrics, uncovering the fact that the classification of corresponding metrics is far from complete and a much richer structure of spacetimes with principal tensor may exist in four and higher dimensions.

Our paper is organized as follows. We start in four dimensions and show that a complete set of Lorentzian metrics admitting the principal tensor, classified long time ago by
Dietz and Rudiger \cite{dietz1981space} and Taxiarchis \cite{Taxiarchis:1985}, can be obtained by a certain limiting procedure starting from the Euclidean off-shell canonical element \eqref{KerrNUTAdSmetric}. Still in four dimensions, Sec.~\ref{sec3} is devoted to a few observations about the Kerr--Schild ansatz in various signatures and its connections to the construction of new canonical metrics. Both  considerations are then extended to higher dimensions in Sec.~\ref{sec4}. Sec.~\ref{sec5} is devoted to conclusions.

\section{Canonical metrics in 4d}\label{sec2}

\subsection{Euclidean signature}

In four dimensions, and denoting by $(x,y,\tau,\psi)=(x_1,x_2,\psi_0, \psi_1)$ and $X=X_1, Y=X_2$, the unique Euclidean metric \eqref{KerrNUTAdSmetric} reads
\ba\label{E4d}
\ts{g}_E&=&\frac{X}{y^2-x^2}(\grad \tau+y^2 \grad \psi)^2+\frac{Y}{x^2-y^2}(\grad \tau+x^2
\grad \psi)^2\nonumber\\
&&+\frac{y^2-x^2}{X}\grad x^2+\frac{x^2-y^2}{Y}\grad y^2\, ,
\ea
where ranges of coordinates $x,y$ must be chosen such that $x^2<y^2$, $X>0$, and $Y<0$.
The principal tensor is determined from
\be\label{Eb4d}
\ts{b}=\frac{1}{2}\bigl[(x^2+y^2)\, \grad \tau +x^2y^2\, \grad \psi\bigr]\, .
\ee
Of course, both the eigenvalues $\{x, y\}$ are spacelike.

The metric is a solution of vacuum Einstein equations, provided we set
\begin{equation}
\begin{aligned}
X&=c_0+c_1x^2+c_2x^4-2b_x x\,,\\
Y&=c_0+c_1y^2+c_2y^4-2b_y y\,,\quad c_2=\Lambda/3\,.
\end{aligned}
\end{equation}
One can check that a suitable choice of coordinate ranges can be achieved by restricting $x$ and $y$ to lie in between the roots of polynomials $X$ and $Y$, respectively.

\subsection{Lorentzian signature}
To obtain the canonical metric \eqref{E4d} in the Lorentzian signature, {additionally to a suitable choice of coordinate ranges, we must} perform the following Wick rotation:
\be\label{Wick}
x=ir\,,\quad \Delta_r=-X\,,\quad \Delta_y=-Y\,,
\ee
to get
\begin{equation}\label{L4d}
\begin{split}
\ts{g}_L&=-\frac{\Delta_r}{\Sigma}(\grad \tau+y^2 \grad \psi)^2+\frac{\Delta_y}{\Sigma}(\grad \tau-r^2
\grad \psi)^2\\
&\quad+\frac{\Sigma}{\Delta_r}\grad r^2+\frac{\Sigma}{\Delta_y}\grad y^2\, ,\quad \Sigma=r^2+y^2\,,
\end{split}
\end{equation}
while the principal tensor reads
\be\label{Lb4d}
\ts{b}=\frac{1}{2}\bigl[(y^2-r^2)\, \grad \tau -r^2y^2\, \grad \psi\bigr]\,.
\ee
Both its eigenvalues $\{r,y\}$ are {non-null}, that is have {non-null} gradients. For a proper choice of coordinate ranges, the angular coordinate $y$ is spacelike, and the causal character of $r$ depends on a sign of the metric function $\Delta_r$, reflecting whether we are below or above the horizon. {$\{\tau,\,\psi\}$ are Killing time and angular coordinates, respectively, with their causal character depending on signs of the metric functions  $\Delta_r$ and  $\Delta_y$.}

{The Kerr--NUT--(A)dS solution of vacuum Einstein equations is recovered upon introducing the} mass $M=-ib_x$ and the NUT parameter $N=b_y$, yielding
\begin{equation}
\begin{aligned}
\Delta_r&=-c_0+c_1r^2-\frac{1}{3}\Lambda r^4-2Mr\,,\\
\Delta_y&=-c_0-c_1y^2-\frac{1}{3}\Lambda y^4+2Ny\,.
\end{aligned}
\end{equation}
{In order to write this solution in the standard form, e.g. \cite{GriffithsPodolsky:2006b}, one has to further introduce the rotation parameter $a$, $c_0=-a^2$, set $c_1=1-a^2\Lambda/3$, and perform the following coordinate transformation to the Boyer--Lindquist coordinates $(t,\phi, r,\theta)$:
\be
y=a\cos\theta\,,\quad \psi=\phi/a\,,\quad \tau=t-a\phi\,.
\ee
}

However, as discovered by
Dietz and Rudiger \cite{dietz1981space} and Taxiarchis \cite{Taxiarchis:1985}, there is yet another off-shell canonical spacetime besides \eqref{L4d}, that admits the 
principal tensor given by \eqref{Lb4d}. It reads\footnote{%
As per usual, in all expressions for the metric we assume a symmetric tensor product in off-diagonal terms.}
\be\label{L4d2}
\ts{g}_{L'}=\frac{\Delta_y}{\Sigma}(\grad \tau-r^2
\grad \psi)^2+\frac{\Sigma}{\Delta_y}\grad y^2+2\grad r(\grad \tau+y^2 \grad \psi)\,.
\ee
The eigenvalue $r$ of the principal tensor is now null {everywhere}, that is
\be
r: \  (\grad r)^2=0\,.
\ee
Despite this fact, the principal tensor is still non-degenerate in the sense of definition in \cite{Houri:2007xz, Krtous:2008tb, Houri:2008th, Houri:2008ng} and generates (in a standard way) both isometries $\ts{\partial}_\tau$ and $\ts{\partial}_\phi$ and a non-trivial Killing tensor in this spacetime.
The metric \eqref{L4d2} becomes a solution of vacuum (with $\Lambda=0)$ Einstein equations provided we set
\be\label{ND}
\Delta_y=2Ny\,.
\ee
We expect it to describe some kind of the NUT charged Aichelberg--Sexl solution.

Although originally derived in a different way \cite{dietz1981space, Taxiarchis:1985}, let us now demonstrate that the new canonical metric \eqref{L4d2} can in fact be obtained by a certain  limit starting from the metric \eqref{L4d}. To this purpose, we rewrite \eqref{L4d} as
\ba\label{KerrtoKSch}
\ts{g}_L&=&-\frac{\Delta_r}{\Sigma}\Bigl((\grad \tau+y^2\grad \psi)^2-\frac{\Sigma^2}{\Delta_r^2}\grad r^2\Bigr)\nonumber\\
&&+\frac{\Delta_y}{\Sigma}(\grad \tau-r^2
\grad \psi)^2+\frac{\Sigma}{\Delta_y}\grad y^2\\
&=&
-\frac{\Delta_r}{\Sigma}\,\ts{l}\,\ts{l}+2\grad r\,\ts{l}+
\frac{\Delta_y}{\Sigma}(\grad \tau-r^2
\grad \psi)^2+\frac{\Sigma}{\Delta_y}\grad y^2\,,\nonumber
\ea
where we introduced a null vector
\be
\ts{l}\equiv\grad \tau+y^2\grad \psi+\frac{\Sigma}{\Delta_r}\grad r\,.
\ee
Defining new coordinates
\be\label{ct1}
\grad \hat \tau=\grad \tau+\frac{r^2}{\Delta_r}\grad r\,,\quad
\grad \hat \psi=\grad \psi+\frac{\grad r}{\Delta_r}\,,
\ee
we find that
\be\label{ll2}
\ts{l}=\grad \hat \tau+y^2\grad \hat \psi\,,
\ee
while $(\grad\tau-r^2\grad\psi)=(\grad\hat \tau-r^2\grad\hat \psi)$. That is, we obtained
\be\label{Lstart}
\ts{g}_L=-\frac{\Delta_r}{\Sigma}\,\ts{l}\,\ts{l}+2\grad r\,\ts{l}+
\frac{\Delta_y}{\Sigma}(\grad \hat \tau-r^2
\grad \hat \psi)^2+\frac{\Sigma}{\Delta_y}\grad y^2\,,
\ee
with $\ts{l}$ given by \eqref{ll2}.
It is now obvious that one can take the limit
\be\label{l1}
\Delta_r\to 0\,,
\ee
the metric remains non-degenerate, while $r$ becomes a null coordinate and we recover the canonical metric \eqref{L4d2}.
One can easily check that neither the coordinate transformation \eqref{ct1} nor the limit \eqref{l1} affect (apart from a gauge term) the potential $\ts{b}$, \eqref{Lb4d}.

Let us note that if one were to perform the limit ${\Delta_r\to 0}$ for the class of on-shell vacuum solutions of Einstein's equations, more care would have to be taken. Namely, the limit ${\Delta_r\to 0}$ effectively sets all the constants, apart from $N$, equal to zero, and influences thus also other metric functions, effectively recovering \eqref{ND}. However, since the metric functions also determine coordinate ranges, these ranges can degenerate in the limit and a suitable rescaling of coordinates has to be performed to resolve such a degeneracy. The limits of this type and the corresponding scaling of coordinates have been recently studied in \cite{Krtous:2015zco}.

\subsection{Other signature}

Let us now perform an additional Wick rotation
\be\label{iz}
y=iz\,,\quad \Delta_z=-\Delta_y\,,\quad \Sigma=r^2-z^2\,.
\ee
The metric \eqref{L4d} then gives
\ba\label{LL4d}
\ts{g}_{LL}&=&-\frac{\Delta_r}{\Sigma}(\grad \tau-z^2 \grad \psi)^2-\frac{\Delta_z}{\Sigma}(\grad \tau-r^2
\grad \psi)^2\nonumber\\
&&+\frac{\Sigma}{\Delta_r}\grad r^2+\frac{\Sigma}{\Delta_z}\grad z^2\,,
\ea
while \eqref{L4d2} yields
\be\label{LL4d2}
\ts{g}_{LL'}=\frac{\Sigma}{\Delta_z}\grad z^2-\frac{\Delta_z}{\Sigma}(\grad \tau-r^2
\grad \psi)^2+2\grad r(\grad \tau-z^2 \grad \psi)\,.
\ee
Both these metrics admit the principal tensor given by
\be\label{LLb4d}
\ts{b}=\frac{1}{2}\bigl[-(z^2+r^2)\, \grad \tau +r^2z^2\, \grad \psi\bigr]\,.
\ee
The only difference is that in the latter case the eigenvalue $r$ is null.

We note that $\ts{g}_{LL}$ is a vacuum solution of the Einstein equations provided
\begin{equation}
\begin{aligned}
\Delta_r&=-c_0+c_1r^2-\frac{1}{3}\Lambda r^4-2Mr\,,\\
\Delta_z&=c_0-c_1z^2+\frac{1}{3}\Lambda z^4-2 Nz\,.
\end{aligned}
\end{equation}
while $\ts{g}_{LL'}$ requires
\be
\Delta_z=-2 Nz\,.
\ee

Let us now apply the procedure above, with the aim to make also the eigenvalue $y$ null. Starting from \eqref{LL4d2}, we can write
\ba
\ts{g}_{LL'}&=&-\frac{\Delta_z}{\Sigma}\Bigl((\grad \tau{-}r^2\grad \psi)^2{-}
   \frac{\Sigma^2}{\Delta_z^2}\grad z^2\Bigr)+2\grad r (\grad \tau{-}z^2\grad \psi)\nonumber\\
&=&
-\frac{\Delta_z}{\Sigma}\,\ts{m}\,\ts{m}+2\grad z\,\ts{m}+2\grad r (\grad \tau-z^2
\grad \psi)\,,
\ea
where we introduced a null vector
\be
\ts{m}\equiv\grad \tau-r^2\grad \psi+\frac{\Sigma}{\Delta_z}\grad z\,.
\ee
Defining new coordinates
\be\label{ct2}
\grad \hat \tau=\grad \tau-\frac{z^2}{\Delta_z}\grad z\,,\quad
\grad \hat \psi=\grad \psi-\frac{\grad z}{\Delta_z}\,,
\ee
we find that
\be\label{mm}
\ts{m}=\grad \hat \tau-r^2\grad \hat \psi\,,
\ee
while $(\grad \tau-z^2
\grad \psi)=(\grad \hat \tau-z^2
\grad \hat \psi)$. This allows one to take the limit
\be\label{l11}
\Delta_z\to 0\,,
\ee
recovering the new metric
\be\label{LL4d3}
\ts{g}_{L'L'}=2\grad z (\grad \hat \tau-r^2\grad \hat \psi)+2\grad r (\grad \hat \tau-z^2
\grad \hat \psi)\,,
\ee
Of course, such a metric is just a flat metric. Nevertheless, it possesses the same non-degenerate principal tensor, given by \eqref{LLb4d}, with now two null eigenvalues:
\be
(\grad r)^2=0\,,\quad (\grad z)^2=0\,.
\ee
Although the corresponding $\ts{h}=\ts{db}$ is reducible (can be written as a skew symmetric product of Killing vectors), it is non-degenerate in the sense of \cite{Houri:2007xz, Krtous:2008tb, Houri:2008th, Houri:2008ng} and gives rise to two Killing vectors and a (reducible) Killing tensor by a standard construction.

To conclude, in this (double minus) signature, there are three possible canonical metrics, \eqref{LL4d}, \eqref{LL4d2}, and \eqref{LL4d3} that all admit the principal tensor
\eqref{LLb4d}.

\section{Variations on the Kerr--Schild form}\label{sec3}

{In this section we obtain the Kerr--Schild form \cite{kerr1965some, Debney:1969zz} of the canonical metric elements. As we shall see,
the above described method for generating the new off-shell canonical metrics shares many features with the procedure for obtaining such a form.}

\subsection{Lorentzian signature}

{We start in the Lorentzian signature, from the Lorentzian metric $\ts{g}_L$ written in the form \eqref{Lstart}. Instead of setting $\Delta_r\to 0$ as we did in the previous section to obtain the metric  \eqref{L4d2}, we now split}
\be
\Delta_r=\tilde \Delta_r-f\,,
\ee
where both $\tilde \Delta_r$ and $f$ are {\em arbitrary} functions of coordinate $r$. By applying the `inverse transformation' to \eqref{ct1} on the tilde part,
\be\label{ct11}
\grad \hat \tau=\grad \tilde \tau+\frac{r^2}{\tilde \Delta_r}\grad r\,,\quad
\grad \hat \psi=\grad \tilde \psi+\frac{\grad r}{\tilde \Delta_r}\,,
\ee
we hence recover the following {\em off-shell Kerr--Schild form} of the canonical metric:
\be\label{KSoffshell}
\ts{g}_L=\ts{\tilde g}_L+\frac{f}{\Sigma}\ts{ll}\,,
\ee
where $\ts{\tilde g}_L$ is the canonical metric \eqref{L4d} with $\Delta_r\to \tilde \Delta_r$ and $\ts{l}$ is a null vector with respect to both $\ts{g}_L$ and $\ts{\tilde g}_L$ metrics:
\ba
\ts{l}&=&\grad \tilde \tau+y^2\grad \tilde \psi+\frac{\Sigma}{\tilde \Delta_r}\grad r\,,\quad \Sigma=r^2+y^2\,,\nonumber\\
\ts{\tilde g}_L&=&-\frac{\tilde \Delta_r}{\Sigma}(\grad \tilde \tau+y^2 \grad \tilde \psi)^2+\frac{\Delta_y}{\Sigma}(\grad \tilde \tau-r^2
\grad \tilde \psi)^2\nonumber\\
&&+\frac{\Sigma}{\tilde \Delta_r}\grad r^2+\frac{\Sigma}{\Delta_y}\grad y^2\,.
\ea
Moreover, the potential $\ts{b}$ reads
\be
\ts{b}=\frac{1}{2}\bigl[(y^2-r^2)\, \grad \tilde \tau -r^2y^2\, \grad \tilde \psi\bigr]\,,
\ee
and generates the principal tensor for both the metric $\ts{g}_L$ and the metric $\ts{\tilde g}_L$. In other words, one can add to the canonical metric $\ts{\tilde g}_L$ a term $\frac{f}{\Sigma}\ts{ll}$, with arbitrary $f(r)$, and the metric still admits the same principal tensor.
We believe that this observation is new and quite interesting.

Note, however, that this is no longer true at the level of Killing tensors. Namely, $\ts{\tilde g}_L$ admits the following Killing tensor:
\begin{align}
    \ts{\tilde k}_L&= \frac{1}{\Sigma}
    \left[ y^2\tilde \Delta_r(\grad \tilde \tau+y^2 \grad \tilde \psi)^2
    +r^2\Delta_y(\grad \tilde \tau-r^2\grad \tilde \psi)^2\right]\nonumber\\
    &
    +\Sigma\left[\frac{r^2 \grad y^2}{\Delta_y}-\frac{y^2\grad r^2}{\tilde \Delta_r}\right]
    \,.\label{k_ccky}
\end{align}
Whereas, the metric $\ts{g}_L$ has
\be
\ts{k}_L=\ts{\tilde k}_L-\frac{f y^2}{\Sigma}\ts{ll}\,.
\ee
Note that such a Killing tensor again takes the `Kerr--Schild form'.

In particular, we may choose
\begin{equation}
\begin{gathered}
f=2Mr\,,\quad \tilde \Delta_r=-c_0+c_1r^2-\frac{1}{3}\Lambda r^4\,,\\
\Delta_y=-c_0-c_1y^2-\frac{1}{3}\Lambda y^4\,,
\end{gathered}
\end{equation}
in which case the metric $\ts{\tilde g}_L$ describes a pure (A)dS space, while the whole metric  $\ts{g}_L$ is the Kerr--(A)dS spacetime, written as a linear in mass deformation of the (A)dS space \cite{kerr1965some, Debney:1969zz}. In this case, $\ts{\tilde k}_L$ is a reducible Killing tensor of the (A)dS space, and becomes a non-reducible Killing tensor of the Kerr--(A)dS spacetime, $\ts{k}_L$, upon adding the $-\frac{f y^2}{\Sigma}\ts{ll}$ term.

\subsection{Other signature}
Similarly, starting from the Lorentzian canonical element \eqref{KSoffshell}, while performing the Wick rotation \eqref{iz}, repeating the steps \eqref{KerrtoKSch}--\eqref{mm}, splitting
\be
\Delta_z=\tilde \Delta_z-g\,,
\ee
and finally applying the `inverse transformation' to \eqref{ct2}, we obtain
the {\em double off-shell Kerr--Schild form}
\be\label{KSoffshell2}
\ts{g}_{LL}=\ts{\tilde g}_{LL}+\frac{f(r)}{\Sigma}\ts{ll}+\frac{g(z)}{\Sigma}\ts{mm}\,,
\ee
where $\Sigma=r^2-z^2\,,$ and
\ba
\ts{l}&=&\grad \tau-z^2\grad \psi+\frac{\Sigma}{\tilde \Delta_r}\grad r\,,\nonumber\\
\ts{m}&=&\grad \tau-r^2\grad \psi+\frac{\Sigma}{\tilde \Delta_z}\grad z\,,\nonumber\\
\ts{\tilde g}_{LL}&=&-\frac{\tilde \Delta_r}{\Sigma}(\grad \tau-z^2 \grad \psi)^2-\frac{\tilde \Delta_z}{\Sigma}(\grad \tau-r^2
\grad \psi)^2\nonumber\\
&&+\frac{\Sigma}{\tilde \Delta_r}\grad r^2+\frac{\Sigma}{\tilde \Delta_z}\grad z^2\,.
\ea
We also have
\be\label{bzzzz}
\ts{b}=\frac{1}{2}\bigl[-(z^2+r^2)\, \grad \tau +r^2z^2\, \grad \psi\bigr]\,.
\ee
for the principal tensor of both  $\ts{g}_{LL}$ and $\ts{\tilde g}_{LL}$.

In particular, choosing
\begin{equation}
\begin{aligned}
f&=2Mr\,,\quad \tilde \Delta_r=-c_0+c_1r^2-\frac{1}{3}\Lambda r^4\,,\\
g&=2\tilde Nz\,,\quad \tilde \Delta_z=c_0-c_1z^2+\frac{1}{3}\Lambda z^4\,,
\end{aligned}
\end{equation}
we recover the special case of the on-shell multi-Kerr--Schild form discussed in all dimensions in \cite{Chen:2007fs}.
Note that the Kerr--NUT--(a)dS metric is written here as a linear in `mass' and linear in `NUT charge' deformation of the (A)dS space.

Another interesting choice is given by
\begin{equation}
\begin{aligned}
f&=\frac{1}{3}\Lambda r^4\,,\quad \tilde \Delta_r=-c_0+c_1r^2-2Mr\,,\\
g&=-\frac{1}{3}\Lambda z^4\,,\quad \tilde \Delta_z=c_0-c_1z^2-2\tilde Nz\,.
\end{aligned}
\end{equation}
This means that the Kerr--NUT--(A)dS metric can also be understood as a linear in $\Lambda$ deformation of the Kerr--NUT metric, perhaps an interesting observation unnoticed in the literature.
Of course, this result may be combined with the above special case, to get Kerr--NUT--(A)dS as a linear deformation of flat space in all: mass, NUT, and $\Lambda$.

Similarly, starting from \eqref{LL4d2}, we arrive at the following off-shell Kerr--Schild metric:
\be
\ts{g}_{LL'}=\ts{\tilde g}_{LL'}+\frac{g(z)}{\Sigma}\ts{mm}\,,
\ee
where $\Sigma=r^2-z^2$ and
\ba
\ts{m}&=&\grad \tau-r^2\grad \psi+\frac{\Sigma}{\tilde \Delta_z}dz\,,\nonumber\\\
\ts{\tilde g}_{LL'}&=&\frac{\Sigma}{\tilde \Delta_z}\grad z^2-\frac{\tilde \Delta_z}{\Sigma}(\grad \tau-r^2
\grad \psi)^2\nonumber\\
&&+2\grad r(\grad \tau-z^2 \grad \psi)\,.
\ea

This metric also admits the principal tensor, given by \eqref{bzzzz}.

{To summarize, we have seen that the procedures for obtaining the new canonical metric element and that for obtaining the Kerr--Schild form are quite similar. Namely,  the key ingredient is to write the original metric in terms of a null vector, and perform a coordinate transformation so that the metric is linear in the associated metric function $\Delta$. The Kerr--Schild form then corresponds to the appropriate split $\Delta=\tilde \Delta-f$, whereas the null limit amounts to setting $\Delta\to 0$.
}

\section{Higher-dimensional generalizations}\label{sec4}

It is obvious that the above presented four-dimensional results can be straightforwardly generalized to higher dimensions. Namely, starting from the Euclidean metric elements \eqref{KerrNUTAdSmetric}--\eqref{Kerreven}, one can perform up to $n$ Wick rotations of the eigenvalues $x_\mu$ and take their `null limit',  to produce a family of new canonical metrics of various signatures. Also the trick for obtaining the off-shell multi-Kerr--Schild form works as expected.

In what follows, let us limit ourselves to the case of Lorentzian signature and write the two
canonical elements and the corresponding off-shell Kerr--Schild form. We first illustrate on a five-dimensional example and then proceed to a general dimension.

\subsection{5d case}
Let us start from the 5-dimensional Euclidean metrics \eqref{Kerrodd}, \eqref{Kerreven}, denoting by $(x,y,\tau,\psi, \phi)=(x_1,x_2,\psi_0, \psi_1,\psi_2)$ and $X=X_1, Y=X_2$. To obtain the Lorentzian version we perform the Wick rotation \eqref{Wick}. This yields
\begin{equation}\label{L15d}
\begin{split}
\ts{g}_L^{(1)}&=-\frac{\Delta_r}{\Sigma}(\grad \tau+y^2 \grad \psi)^2+\frac{\Delta_y}{\Sigma}(\grad \tau-r^2
\grad \psi)^2\\
&\quad+\frac{\Sigma}{\Delta_r}\grad r^2+\frac{\Sigma}{\Delta_y}\grad y^2\\
&\quad-\frac{c}{r^2y^2}\bigl(\grad \tau+(y^2-r^2)\grad \psi-r^2y^2 \grad \phi\bigr)^2\,,
\end{split}
\end{equation}
respectively,
\begin{equation}\label{L25d}
\begin{split}
\ts{g}_L^{(2)}&=-\frac{\Delta_r}{\Sigma}(\grad \tau+y^2 \grad \psi)^2+\frac{\Delta_y}{\Sigma}(\grad \tau-r^2
\grad \psi)^2\\
&\quad+\frac{\Sigma}{\Delta_r}\grad r^2+\frac{\Sigma}{\Delta_y}\grad y^2-{c}{r^2y^2}\grad \phi^2\,,
\end{split}
\end{equation}
where $\Sigma=r^2+y^2$, and assuming $c<0$, $\Delta_y>0$. The principal tensor is given by
\be\label{Lb5d}
\ts{b}=\frac{1}{2}\bigl[(y^2-r^2)\, \grad \tau -r^2y^2\, \grad \psi\bigr]\,,
\ee
both its eigenvalues $\{r,y\}$ are spacelike.

The metric becomes a vacuum solution provided we set
\begin{equation}
\begin{split}
\Delta_r^{(1)}&=-\frac{c}{r^2}+c_1r^2-\frac{1}{6}\Lambda r^4+2M\,,\\
\Delta_y^{(1)}&=\frac{c}{y^2}-c_1y^2-\frac{1}{6}\Lambda y^4+2N\,,
\end{split}
\end{equation}
respectively
\begin{equation}
\begin{split}
\Delta_r^{(2)}&=c_1r^2-\frac{1}{6}\Lambda r^4+2M\,,\\
\Delta_y^{(2)}&=c_1y^2-\frac{1}{6}\Lambda y^4+2N\,,
\end{split}
\end{equation}
where we have denoted $M=b_x$ and $N=b_y$.

Note that
\be
\ts{g}_L^{(1,2)}+\frac{f(r)}{\Sigma}\ts{ll}\,,\quad \ts{l}=\grad \tau+y^2\grad \psi+\frac{\Sigma}{\Delta_r}\grad r\,,
\ee
{where $f(r)$ is an arbitrary function,}
are again the off-shell Kerr--Schild metrics with the same principal tensor \eqref{Lb5d}.

{By performing the transformation
\begin{equation}
\grad \tau=\grad \hat \tau-\frac{r^2}{\Delta_r}\grad r\,,\quad
\grad \psi=\grad \hat \psi-\frac{\grad r}{\Delta_r}\,,
\end{equation}
complemented with
\begin{equation}
\grad \phi=\grad \hat \phi-\frac{1}{r^2\Delta_r}\grad r\,,
\end{equation}
for the metric \eqref{L15d} and
\begin{equation}
\grad \phi=\grad \hat \phi\,,
\end{equation}
for the metric \eqref{L25d}, we can take the limit $\Delta_r\to 0$ to obtain}
\begin{align}
\begin{split}\label{LL15d}
{\ts{g}_{L'}^{(1)}}&=\frac{\Delta_y}{\Sigma}(\grad \hat\tau-r^2\grad \hat\psi)^2+\frac{\Sigma}{\Delta_y}\grad y^2
+2\grad r (\grad \hat\tau+y^2 \grad \hat\psi)\\
&\quad-\frac{c}{r^2y^2}\bigl(\grad \hat\tau+(y^2-r^2)\grad \hat\psi-r^2y^2 \grad \hat\phi\bigr)^2\,,
\end{split}\raisetag{6ex}\\
\begin{split}\label{LL25d}
{\ts{g}_{L'}^{(2)}}&=\frac{\Delta_y}{\Sigma}(\grad \hat\tau-r^2\grad \hat\psi)^2+\frac{\Sigma}{\Delta_y}\grad y^2\\
&\quad+2\grad r (\grad \hat\tau+y^2 \grad \hat\psi)-{c}r^2y^2 \grad \hat\phi^2\,,
\end{split}
\end{align}
respectively, both admitting the same principal tensor \eqref{Lb5d}. The first metric, $\ts{g}_L^{(1)}$ does not solve the vacuum equations for any choice of $\Delta^{(1)}_y$, whereas the second one is a solution provided we set
\be
\Delta_y^{(2)}=2N\,.
\ee

\subsection{General dimension}

To write the Lorentzian canonical metrics in a general dimension, we perform the following Wick rotation:
\be
x_\dg=ir\,,\quad X_\dg=-\Delta\,,\quad U_\dg=\Sigma\,,
\ee
leaving all other $X_\mu$'s (contrary to the previous sections) unchanged.
We also understand all functions having $x_\dg$-dependence replaced by $ir$, for example,
\be
\Sigma=\prod_{\nu=1}^{n-1}(x_{\nu}^2+r^2)\,.
\ee
The Lorentzian canonical element in even dimensions, \eqref{KerrNUTAdSmetric}, then reads
\begin{equation}\label{LevenHD}
\begin{split}
\tens{g}_{L\,2n}
  &=-\frac{\Delta}{\Sigma}\Bigl(\,\sum_{j=0}^{\dg-1} \A{j}_{\dg}\grad\psi_j \Bigr)^{\!2}+\frac{\Sigma}{\Delta}\grad r^2\\
  &\quad+
   \sum_{\mu=1}^{\dg-1}\;\biggl[\; \frac{U_\mu}{X_\mu}\,{\grad x_{\mu}^{2}}
  +\, \frac{X_\mu}{U_\mu}\,\Bigl(\,\sum_{j=0}^{\dg-1} \A{j}_{\mu}\grad\psi_j \Bigr)^{\!2}
  \;\biggr]\,.\quad
\end{split}\raisetag{10ex}
\end{equation}
In odd dimensions, corresponding to  \eqref{Kerrodd} and \eqref{Kerreven}, we have
\begin{align}
\tens{g}_{L\,2n+1}^{(1)}&=\tens{g}_{L\,2n}
  +\frac{c}{\A{\dg}}\Bigl(\sum_{k=0}^\dg \A{k}\grad\psi_k\!\Bigr)^{\!2}\;,\label{Lodd1HD}\\
\tens{g}_{L\,2n+1}^{(2)}&=\tens{g}_{L\,2n}+c\A{n}\grad \psi_n^2\,.\label{Lodd2HD}
\end{align}
All of them admit the principal tensor given by \eqref{PCCKYpot},
\begin{equation}\label{PCCKYpotFinal}
\ts{b} = \frac12 \sum_{k=0}^{\dg-1}\A{k+1}\grad\psi_k\;.
\end{equation}
The metrics become vacuum solutions provided we set
\begin{equation}
\begin{aligned}
\text{even $D$\,:}\quad &\Delta=-\sum_{k=0}^n c_k (-r^2)^{k}-2Mr\,,\\
\text{odd $D$\,:}\quad  &\Delta^{(1)}=-\sum_{k=1}^n c_k (-r^2)^{k}+2M-\frac{c}{r^2}\,,\\
                        &\Delta^{(2)}=-\sum_{k=1}^n c_k (-r^2)^{k}+2M\,,
\end{aligned}
\end{equation}
while other $X_\mu$'s ($\mu=1,\dots,n-1)$ are given in \eqref{Xmus}.

As in the lower-dimensional cases, we introduce a null vector
\begin{equation}\label{lambdadef}
\ts{l}=\sum_{j=0}^{\dg-1} \A{j}_{\dg}\grad\psi_j+\frac{\Sigma}{\Delta}\grad r\,,
\end{equation}
and change coordinates as
\begin{equation}\label{hatpsiHD}
    \grad\hat\psi_j = \grad\psi_j + \frac{r^{2(\dg{-}1{-}j)}}{\Delta}\grad r\;,
\end{equation}
with the expression for $\hat\psi_\dg$ modified to just ${\hat\psi_\dg=\psi_\dg}$ in the second odd-dimensional case \eqref{Lodd2HD}. Here, ${\hat\tau\equiv\hat\psi_0}$ plays a role of time coordinate. Such a change satisfies\;
${\sum_{j=0}^{\dg-1} \A{j}_{\mu}\grad\psi_j = \sum_{j=0}^{\dg-1} \A{j}_{\mu}\grad\hat\psi_j}$\; for ${\mu=1,\dots,\dg-1}$,
${\sum_{k=0}^{\dg} \A{k}\grad\psi_k = \sum_{k=0}^{\dg} \A{k}\grad\hat\psi_k}$, and $\ts{l}$ simplifies to
\begin{equation}\label{lambdadef}
\ts{l}=\sum_{j=0}^{\dg-1} \A{j}_{\dg}\grad\hat\psi_j\,.
\end{equation}
Here we have extensively used the identity
\begin{equation}\label{Axid}
    \sum_{j=0}^{\dg-1}\A{j}_\mu \frac{(-x_\nu^2)^{\dg{-}1{-}j}}{U_\nu} = \delta^\nu_\mu\;.
\end{equation}
The metric takes the form
\begin{equation}\label{LevenHDll}
\begin{split}
\tens{g}_{L\,2n}
  &=-\frac{\Delta}{\Sigma}\ts{ll}+2\ts{l}\,\grad r\\
  &\quad+
   \sum_{\mu=1}^{\dg-1}\;\biggl[\; \frac{U_\mu}{X_\mu}\,{\grad x_{\mu}^{2}}
  +\, \frac{X_\mu}{U_\mu}\,\Bigl(\,\sum_{j=0}^{\dg-1} \A{j}_{\mu}\grad\hat\psi_j \Bigr)^{\!2}
  \;\biggr]\,,
\end{split}\raisetag{10ex}
\end{equation}
supplemented by \eqref{Lodd1HD} or \eqref{Lodd2HD} with $\hat\psi_j$ instead of $\psi_j$ in the odd-dimensional case.

Clearly, splitting the metric function $\Delta=\tilde{\Delta}-f(r)$ yields the off-shell Kerr--Schild form
\be
\ts{g}_{L}=\tilde{\ts{g}}_L+\frac{f(r)}{\Sigma}\ts{ll}\,,
\ee
where
{$\tilde{\ts{g}}_{L}$ is given by \eqref{LevenHD}--\eqref{Lodd2HD} with $\Delta$ replaced by $\tilde \Delta$.
Both $\ts{g}_{L}$} and $\tilde{\ts{g}}_{L}$ admit the same principal tensor. (If other signatures were considered, we would recover the off-shell multi-Kerr--Schild form, generalizing the results in  \cite{Chen:2007fs}.)

In the metric \eqref{LevenHDll} we can easily perform the limit ${\Delta\to0}$, making the eigenvalue $r$ null, and obtaining the new canonical elements:
\begin{equation}\label{HD1}
\begin{split}
\tens{g}_{L'\,2n}
  &=2\grad r \Bigl(\,\sum_{j=0}^{\dg-1} \A{j}_{\dg}\grad\hat\psi_j \Bigr)\\
  &\quad+
   \sum_{\mu=1}^{\dg-1}\;\biggl[\; \frac{U_\mu}{X_\mu}\,{\grad x_{\mu}^{2}}
  +\, \frac{X_\mu}{U_\mu}\,\Bigl(\,\sum_{j=0}^{\dg-1} \A{j}_{\mu}\grad\hat\psi_j \Bigr)^{\!2}
  \;\biggr]\,,\quad
\end{split}\raisetag{10ex}
\end{equation}
and
\begin{align}
\tens{g}_{L'\,2n+1}^{(1)}&=\tens{\tilde g}_{L\,2n}+\frac{c}{\A{\dg}}\Bigl(\sum_{k=0}^\dg \A{k}\grad\hat\psi_k\!\Bigr)^{\!2}\;,\\
\tens{g}_{L'\,2n+1}^{(2)}&=\tens{\tilde g}_{L\,2n}+c\A{n}\grad \hat\psi_n^2\,.\label{HD3}
\end{align}
These metrics become the vacuum solutions of Einstein equations (with $\Lambda=0$) provided we set
\begin{equation}\label{Xmus2}
\begin{aligned}
\text{even $D$\,:} &\quad X_\mu=-2b_\mu x_\mu\,,\\
\text{odd $D$\,:}  &\quad X_\mu^{(2)}=-2b_\mu x_\mu\,,
\end{aligned}
\end{equation}
whereas there is no solution for $X_\mu^{(1)}$.

Let us finally note that by introducing the following veilbein {($\mu=1,\dots,\dg-1)$:
\begin{equation}\label{Darbouxformfr}
\begin{gathered}
\ts{k}=\grad r\,,\quad \ts{l}=\sum_{j=0}^{\dg-1} \A{j}_{1}\grad\hat\psi_j\,,\\
\enf\mu = {\Bigl(\frac{U_\mu}{X_\mu}\Bigr)^{\!\frac12}}\grad x_{\mu}\;,\quad
\ehf\mu = {\Bigl(\frac{X_\mu}{U_\mu}\Bigr)^{\!\frac12}}
  \sum_{j=0}^{\dg-1}\A{j}_{\mu}\grad\hat\psi_j\;,\\
\ezf_{(1)} = {\Bigl(\frac{c}{\A{\dg}}\Bigr)^{\frac12}}\sum_{k=0}^{\dg}\A{k}\grad\hat\psi_k\,,\;\,
\ezf_{(2)}=\sqrt{c\A{n}}\grad \hat\psi_n\,,
\end{gathered}
\end{equation}
where $\ts{k}$ and $\ts{l}$ are null, with $\ts{k}\cdot \ts{l}=1$},
the new metrics \eqref{HD1}--\eqref{HD3}, together with their principal tensor can be written as
\begin{gather}
\ts{h} = -r\,\ts{k}\wedge \ts{l}+\sum_{\mu=1}^{\dg-1} x_\mu \enf{\mu}\wedge\ehf{\mu}\,,\\
\ts{g}_{L'} = 2\ts{kl}+\sum_{\mu=2}^n \bigl(
      \enf\mu\,\enf\mu + \ehf\mu\,\ehf\mu
      \bigr) + \eps\, \ezf\,\ezf\,,\label{orthonormmtrc2}
\end{gather}
which is a `null Lorentzian version' of the Darboux frame~\eqref{orthonormmtrc}.

\section{Conclusions}\label{sec5}
It has been believed that the classification of metrics admitting the principal tensor, that is a {\em non-degenerate} closed conformal Killing--Yano 2-form, is completed
and uniquely leads to the off-shell Kerr--NUT--(A)dS metric (reviewed in the Introduction). However, as shown in this paper, this is not true unless the signature of the metric is  Euclidean. The original classification assumed the Euclidean signature for the metric and the canonical Darboux form for the
principal tensor, see Eqs.~\eqref{orthonormmtrc}. Consequently, also the eigenvalues of the principal tensor were assumed to have spacelike character.

In this paper we have shown how to construct new canonical elements whose characteristic feature is that one or more of the eigenvalues of the principal tensor are null.
Among these, the new Lorentzian canonical elements \eqref{HD1}--\eqref{HD3} are perhaps of the biggest interest, generalizing the 4-dimensional metric element constructed in \cite{dietz1981space, Taxiarchis:1985}. These metrics can formally be constructed by a procedure similar to obtaining the Kerr--Schild form. Namely, one introduces a null vector and performs a coordinate transformation such that the metric becomes linear in the corresponding metric function. Such a metric function can then be sent to zero, making the associated  eigenvalue null, while other functions remain unspecified. Going back to the canonical coordinates recovers the new metric elements \eqref{HD1}--\eqref{HD3}. This is a clever trick
for how to switch off one of the metric functions without making the metric singular. In this sense, the new metrics can be considered as a `special case' of the off-shell Kerr--NUT--(A)dS metric.

The presented results re-open the problem of classifying the metrics admitting the principal tensor. Of special importance to physics are of course the metrics with Lorentzian signature.
While in this paper we uncovered some new such metrics, the classification is far for complete. In this paper we simply concentrated on the metrics for which the principal tensor takes the `null Lorentzian Darboux form' \eqref{orthonormmtrc2}, with $r$ a null eigenvalue. However, in the Lorentzian signature there is many more possibilities for the canonic form of a non-degenerate 2-form, see e.g. \cite{Milson:2004wr}. For this reason the problem of classifying all Lorentzian metrics admitting the principal tensor still remains open and will be discussed elsewhere \cite{Frolov:prep}.

\section*{Acknowledgments}

We would like to thank the anonymous referee for helping us to improve this manuscript.
V.F.\ thanks the Natural Sciences and Engineering Research Council of Canada and the Killam Trust for financial support.
P.K.\ is supported by the project of excellence of the Czech Science Foundation \mbox{No.~14-37086G}.
D.K.\ is supported by the Perimeter Institute for Theoretical Physics and by the Natural Sciences and Engineering Research Council of Canada. Research at Perimeter Institute is supported by the Government of Canada through
the Department of Innovation, Science and Economic Development Canada and by
the Province of Ontario through the Ministry of Research, Innovation and Science.


\providecommand{\href}[2]{#2}\begingroup\raggedright\endgroup

\end{document}